\documentclass[aps,twocolumn,prl,tightenlines,floatfix,nopacs]{revtex4}
\usepackage[dvips]{graphicx}
\usepackage{amsmath}
\usepackage{amssymb}
\usepackage{times}
\usepackage[english]{babel}

\begin{document}

\textbf{Comment on ``Density and Spin response of a strongly-interacting
Fermi gas in the attractive and quasi-repulsive regime [Phys. Rev. 108, 080401 (2012)]"}

In Ref.\cite{StrinatiPRL} the authors
summarize a linear response theory for Fermi gases undergoing BCS-BEC crossover. Invoking
Popov theory, they include a
rather complex set of diagrams on the basis of avoiding a
divergence in the density response associated with non-interacting
bosons.
We wish to point out that
repairing this divergence
as they do above $T_c$  
leads to violations of Ward identities or
conservation principles, a manifestation of which
is an unphysical Meissner
effect in the normal state. 
Indeed, their previous work \cite{PS05}
implied that consistency was not yet established in
their Popov-based approach;
the authors indicated
that in the last step of the program they would need to ``modify the
number equation to be consistent with the above approximation"
for their self energy diagrams.
Although they did not specify the precise form of the
self energy $\Sigma_p$, the Supplementary Material in Ref.~\cite{StrinatiPRL}
suggested that there were no incompatibilities between the 
linear response vertex $\Lambda$
and 
$\Sigma_p$.
Here we prove that this is not the case.

We confine our attention in this Comment to the attractive
interaction regime which was considered by us in earlier work
\cite{OurAnnPhys,HaoPRL1,HaoPRL2,NJOP,Ourconduct} to address
a systematic theory of spin and charge linear response
above and below $T_c$. Our work, not cited by the authors, was
shown to be demonstrably consistent with a number of sum rules,
reflecting gauge invariance.
In
Appendix D of Ref.~\cite{OurAnnPhys} we presented the appropriate set
of diagrams for the vertex $\Lambda$ 
which is compatible with the self energy $\Sigma$ 
associated with the Nozieres Schmitt-Rink approach. Here
the t-matrix contains only bare Green's functions.
Adding additional diagrams in any transport theory has to be done with considerable
care to maintain the Ward identity between
$\Lambda$
and $\Sigma$.

The diagrams of Ref.~\cite{StrinatiPRL} involve the
Maki-Thompson (MT) and Aslamazov-Larkin (AL) diagrams, which as we
showed earlier 
\cite{OurAnnPhys}
can be made compatible
with a Ward identity;
however, in addition to this subset they introduce an infinite set
of diagrams of the AL type.  The fermionic self energy
($\Sigma_P$, to which they refer in the Supplemental Material)
is not available for explicit checking of the Ward identity: 
$ Q\cdot\Lambda(K,K^+)=G^{-1}(K)-G^{-1}(K^+)$ with $K^+=K+Q$. 
Nevertheless 
it is possible to prove that, 
by reconstructing additional self energy contributions from the new AL diagrams, there must be additional diagrams in the response vertex as well.
In particular we find a violation of gauge invariance in the form 
\begin{eqnarray}
& &Q\cdot\big[\lambda(K,K^+)+\textrm{MT}(K,K^+)+\textrm{AL}(K,K^+)\nonumber\\
&+&\sum_{n=2}^{\infty}\textrm{AL}_n(K,K^+)\big]
\neq G^{-1}(K)-G^{-1}(K^+).
\end{eqnarray}
The reason that this identity cannot be respected is because the repeated 
AL diagrams will introduce a factor of $2^{n}$ in the response vertex due to the two possible ways of 
connecting the two fermion propagators inside.  In order to find consistency
it is essential to include diagrams of the MT or mixed MT and AL
form and thereby cancel these
factors of $2^{n}$ which do not appear naturally in self energy diagrams.

Moreover, we object to the authors' statement in
Ref.~\cite{StrinatiPRL} claiming that ``pairing fluctuations beyond mean
field can be included below $T_c$
following Ref.~\cite{StrinatiPRB70}".
In the broken-symmetry phase it has long been known that a consistent calculation
of the compressibility based on the longitudinal density-density correlation functions
will require that the
collective modes of the order parameter are also included, 
as we have discussed elsewhere \cite{HaoPRL1}.
Importantly,
they are very complicated to address away from the BCS regime.
Figure 3 from Ref.~\cite{StrinatiPRB70}
also illustrates a problematic first order transition, which is known to be intrinsic to
Popov theory \cite{ShiGriffin} and will lead to an ill-behaved compressibility below $T_c$.

The burden of proving consistency with a Ward identity lies with
the authors, not the readership. 
No theory of transport in a many body system
can be shown to be correct until this level of consistency is proved. When there
is such a violation, f-sum rules will not be satisfied and a normal state
Meissner effect ensues.
This Comment is intended to
emphasize
the key role played by
sum rules and other conservation
constraints in transport and scattering theories of Fermi gases.
The quantitative agreement with experiments
reported in Ref. \cite{StrinatiPRL} is not a valid reason
for ignoring theoretical inconsistencies.

\rm{
Chih-Chun Chien$^{1}$, Hao Guo$^{2}$, and K. Levin$^{3}$}

$^{1}$Theoretical Division, Los Alamos National Laboratory, MS B213, Los Alamos, NM 87545, USA
\vskip1mm
$^{2}$University of Hong Kong, Hong Kong, China
\vskip1mm
$^{3}$James Frank Institute and department of physics, University of Chicago, Chicago,
IL 60637, USA

\bibliographystyle{apsrev}

\begin{thebibliography}{9}
\expandafter\ifx\csname natexlab\endcsname\relax\def\natexlab#1{#1}\fi
\expandafter\ifx\csname bibnamefont\endcsname\relax
  \def\bibnamefont#1{#1}\fi
\expandafter\ifx\csname bibfnamefont\endcsname\relax
  \def\bibfnamefont#1{#1}\fi
\expandafter\ifx\csname citenamefont\endcsname\relax
  \def\citenamefont#1{#1}\fi
\expandafter\ifx\csname url\endcsname\relax
  \def\url#1{\texttt{#1}}\fi
\expandafter\ifx\csname urlprefix\endcsname\relax\def\urlprefix{URL }\fi
\providecommand{\bibinfo}[2]{#2}
\providecommand{\eprint}[2][]{\url{#2}}

\bibitem[{\citenamefont{Palestini et~al.}(2012)\citenamefont{Palestini, Pieri,
  and Strinati}}]{StrinatiPRL}
\bibinfo{author}{\bibfnamefont{F.}~\bibnamefont{Palestini}},
  \bibinfo{author}{\bibfnamefont{P.}~\bibnamefont{Pieri}}, \bibnamefont{and}
  \bibinfo{author}{\bibfnamefont{G.~C.} \bibnamefont{Strinati}},
  \bibinfo{journal}{Phys. Rev. Lett.} \textbf{\bibinfo{volume}{108}},
  \bibinfo{pages}{080401} (\bibinfo{year}{2012}).

\bibitem[{\citenamefont{Pieri and Strinati}(2005)}]{PS05}
\bibinfo{author}{\bibfnamefont{P.}~\bibnamefont{Pieri}} \bibnamefont{and}
  \bibinfo{author}{\bibfnamefont{G.~C.} \bibnamefont{Strinati}},
  \bibinfo{journal}{Phys. Rev. B} \textbf{\bibinfo{volume}{71}},
  \bibinfo{pages}{094520} (\bibinfo{year}{2005}).

\bibitem[{\citenamefont{Levin et~al.}(2010)\citenamefont{Levin, Chen, Chien,
  and He}}]{OurAnnPhys}
\bibinfo{author}{\bibfnamefont{K.}~\bibnamefont{Levin}},
  \bibinfo{author}{\bibfnamefont{Q.~J.} \bibnamefont{Chen}},
  \bibinfo{author}{\bibfnamefont{C.~C.} \bibnamefont{Chien}}, \bibnamefont{and}
  \bibinfo{author}{\bibfnamefont{Y.}~\bibnamefont{He}}, \bibinfo{journal}{Ann.
  Phys.} \textbf{\bibinfo{volume}{325}}, \bibinfo{pages}{233}
  (\bibinfo{year}{2010}).

\bibitem[{\citenamefont{Guo et~al.}(2010)\citenamefont{Guo, Chien, and
  Levin}}]{HaoPRL1}
\bibinfo{author}{\bibfnamefont{H.}~\bibnamefont{Guo}},
  \bibinfo{author}{\bibfnamefont{C.~C.} \bibnamefont{Chien}}, \bibnamefont{and}
  \bibinfo{author}{\bibfnamefont{K.}~\bibnamefont{Levin}},
  \bibinfo{journal}{Phys. Rev. Lett.} \textbf{\bibinfo{volume}{105}},
  \bibinfo{pages}{120401} (\bibinfo{year}{2010}).

\bibitem[{\citenamefont{Guo et~al.}(2011{\natexlab{a}})\citenamefont{Guo,
  Wulin, Chien, and Levin}}]{HaoPRL2}
\bibinfo{author}{\bibfnamefont{H.}~\bibnamefont{Guo}},
  \bibinfo{author}{\bibfnamefont{D.}~\bibnamefont{Wulin}},
  \bibinfo{author}{\bibfnamefont{C.~C.} \bibnamefont{Chien}}, \bibnamefont{and}
  \bibinfo{author}{\bibfnamefont{K.}~\bibnamefont{Levin}},
  \bibinfo{journal}{Phys. Rev. Lett.} \textbf{\bibinfo{volume}{107}},
  \bibinfo{pages}{020403} (\bibinfo{year}{2011}{\natexlab{a}}).

\bibitem[{\citenamefont{Guo et~al.}(2011{\natexlab{b}})\citenamefont{Guo,
  Wulin, Chien, and Levin}}]{NJOP}
\bibinfo{author}{\bibfnamefont{H.}~\bibnamefont{Guo}},
  \bibinfo{author}{\bibfnamefont{D.}~\bibnamefont{Wulin}},
  \bibinfo{author}{\bibfnamefont{C.~C.} \bibnamefont{Chien}}, \bibnamefont{and}
  \bibinfo{author}{\bibfnamefont{K.}~\bibnamefont{Levin}},
  \bibinfo{journal}{New J. Phys.} \textbf{\bibinfo{volume}{13}},
  \bibinfo{pages}{075011} (\bibinfo{year}{2011}{\natexlab{b}}).

\bibitem[{\citenamefont{Wulin et~al.}(2011)\citenamefont{Wulin, M~Fregoso, Guo,
  Chien, and Levin}}]{Ourconduct}
\bibinfo{author}{\bibfnamefont{D.}~\bibnamefont{Wulin}},
  \bibinfo{author}{\bibfnamefont{B.}~\bibnamefont{M~Fregoso}},
  \bibinfo{author}{\bibfnamefont{H.}~\bibnamefont{Guo}},
  \bibinfo{author}{\bibfnamefont{C.-C.} \bibnamefont{Chien}}, \bibnamefont{and}
  \bibinfo{author}{\bibfnamefont{K.}~\bibnamefont{Levin}},
  \bibinfo{journal}{Phys. Rev. B} \textbf{\bibinfo{volume}{84}},
  \bibinfo{pages}{140509(R)} (\bibinfo{year}{2011}).

\bibitem[{\citenamefont{Pieri et~al.}(2004)\citenamefont{Pieri, Pisani, and
  Strinati}}]{StrinatiPRB70}
\bibinfo{author}{\bibfnamefont{P.}~\bibnamefont{Pieri}},
  \bibinfo{author}{\bibfnamefont{L.}~\bibnamefont{Pisani}}, \bibnamefont{and}
  \bibinfo{author}{\bibfnamefont{G.~C.} \bibnamefont{Strinati}},
  \bibinfo{journal}{Phys. Rev. B} \textbf{\bibinfo{volume}{70}},
  \bibinfo{pages}{094508} (\bibinfo{year}{2004}).

\bibitem[{\citenamefont{Shi and Griffin}(1998)}]{ShiGriffin}
\bibinfo{author}{\bibfnamefont{H.}~\bibnamefont{Shi}} \bibnamefont{and}
  \bibinfo{author}{\bibfnamefont{A.}~\bibnamefont{Griffin}},
  \bibinfo{journal}{Phys. Rep.} \textbf{\bibinfo{volume}{304}},
  \bibinfo{pages}{1} (\bibinfo{year}{1998}).

\end{thebibliography}

\end{document}